# Constructive Axiomatics in Spacetime Physics Part II: Constructive Axiomatics in Context

Emily Adlam,* Niels Linnemann† and James Read‡

**Abstract**

The Ehlers-Pirani-Schild (EPS) constructive axiomatisation of general relativity, published in 1972, purports to build up the kinematical structure of that theory from only axioms which have indubitable empirical content. It is, therefore, of profound significance both to the epistemology and to the metaphysics of spacetime theories. In this article, we set the EPS approach in its proper context, by (a) discussing the history of constructive approaches to spacetime theories in the lead-up to EPS; (b) addressing some of the major concerns raised against EPS; (c) considering how EPS compares with 'chronometric' approaches to affording the metric field of general relativity its operational significance; (d) distinguishing quite generally between different kinds of constructive approach, and fitting EPS into this classification; (e) discussing how constructivism bears on a number of other issues in the foundations of physics; and (f) assessing the merits of constructivism *qua* local foundationalist project. There are two companion papers, in which we provide a pedagogical walkthrough to the EPS axiomatisation (Part I), and discuss/develop versions of EPS with quantum mechanical inputs (Part III).

## Contents

*eadlam2@uwo.ca
†niels.linnemann@uni-bremen.de
‡james.read@philosophy.ox.ac.uk

# 1 Introduction

## 1.1 Prehistory: Reichenbach on constructive axiomatics

In his 1924 book, *Axiomatization of the Theory of Relativity*, Hans Reichenbach proposes a method of *constructive axiomatics*: physical theories, such as Einstein's theory of relativity, should be formulated in such a way that their content is built up from axioms which have direct and indubitable empirical content; in this way, the physical content of the theory under consideration is (supposedly) rendered manifest (Reichenbach, 1969). Here is Reichenbach himself on the idea:

> It is possible to start with the observable facts and to end with the abstract conceptualization. A certain loss in formal elegance will be balanced by logical clarity. The empirical character of the axioms is immediately evident, and it is easy to see what consequences follow from their respective confirmations and disconfirmations. Such a *constructive* axiomatization is more in line with physics than a *deductive* one [that runs from the abstract to the observable], because it serves to carry out the primary aim of physics, the description of the physical world. (Reichenbach, 1969, p. 5)

The idea of constructive axiomatics has since found its way into the broader literature on the foundations of physics. Here is, for instance, Carrier on constructive axiomatics with a similar characterisation:

In a constructive axiomatization only those statements are accepted as fundamental that are immediately amenable to experimental control. A deductive axiomatization, by contrast, confers fundamental status to more abstract statements (such as variational principles). ((Carrier, 1990, p. 370)

It is important to clarify at this early point that a constructive axiomatisation need not be explicitly formulated through constructive vocabulary. Rather, any term can be used that is linearly defined through constructive terms or through terms that have already been linearly defined through constructive terms. So, to stress: constructive axiomatisation only employs constructive terms *or* terms in line with a postulate of semantic linearity (Carrier, 1990)—but not just constructive terms. We refer to any axiomatisation that follows semantic linearity—independently of whether the basic vocabulary used is constructive or not—'constructivist'; the noun constructivism will only be used with respect to this adjective 'constructivist'.

Now, Reichenbach's attempt to axiomatise constructively the theory of relativity was not wholly successful. In the context of the special theory, for instance, he failed to appreciate that the paths of light rays—the motions of which he used as his empirically-motivated constructive axioms—do not in themselves suffice to recover the theory: one needs, in addition, either (i) projective structure, which associated with the paths of massive particles, or (ii) certain topological assumptions (regarding the compactification of $\mathbb{R}^4$ by a light cone at conformal infinity—for the details and history, see Rynasiewicz (2005)). Such issues were pointed out repeatedly to Reichenbach by Weyl (who would go on to present his own constructive approach to relativity theory), but were—much to Weyl's chagrin—seemingly ignored.[1] In the context of the general theory, Reichenbach sought to use idealised rods and clocks in local neighbourhoods to recover the metric field: but one might reasonably complain that (i) his use of 'local neighbourhoods' is not sufficiently mathematically well-defined,[2] and (ii) in a constructive approach, one cannot *presuppose* the existence of such complex bodies as rods and clocks.[3] Again, both of these issues were (seemingly) overcome by Weyl, who was able to show that a given Lorentzian metric field is fixed (up to a spacetime-dependent scale factor) by its projective structure (associated with the paths of massive bodies) and conformal structure (associated with the paths of massless bodies) without any need to make recourse to clocks or rods (Weyl, 1921). However—for reasons which we will explain in what follows—it would in fact be fifty years until a constructive axiomatisation of general relativity in the spirit of Weyl was articulated fully.

---

[1] Again, see Rynasiewicz (2005) for a fascinating history of this episode.

[2] Such concerns persist a century later—see e.g. Weatherall (2021) in response to Read et al. (2018).

[3] Famously, this was one of the later Einstein's misgivings about the theory of special relativity as presented in his 1905 paper (Einstein, 1905)—for the history here, see (Giovanelli, 2014). In 1919, Einstein identified his 1905 formulation of special relativity as a 'principle theory'—which is a theory built from observed empirical regularities, raised to the status of postulates (Einstein, 1919). There are clear affinities between principle theories and constructive axiomatisations—discussed in detail later in this article—so it is of little surprise that the 'cannot presuppose complex bodies' complaint should arise in both cases.

## 1.2 The Ehlers-Pirani-Schild axiomatisation

The project of the constructive axiomatisation of general relativity can be seen as being taken up once again in 1972 by Ehlers, Pirani and Schild (henceforth EPS), in an attempt to counter the by-then dominant *deductive* chronometric method. On the chronometric method, as championed by e.g. Synge (1959; 1960), good clocks are *stipulated* to read off the worldline interval of the path along which they travel.[4] The approach is deductive as testable hypotheses—ultimately statements on the worldline intervals of a path—are derived from the *readily presupposed* theory of general relativity.

In contrast to this methodology, EPS took up Weyl's above-mentioned result in an attempt to complete the constructive project. *Prima facie*, no clock is needed on this approach to establishing the metric structure of spacetime, because Weyl's theorem is understood to provide a direct way to link much of the theoretical structure of general relativity to local experience (namely via the allegedly relatively theory-neutral notions of free particles and light rays) without the recourse to theory-external proxies such as clocks, as already used in the deductive approach.[5]

Building on Weyl's work, EPS subsequently managed to show that the causal-inertial method can even serve as a suitable basis for a constructive axiomatisation: all relevant structure from the differentiable manifold up to the metric was more-or-less rigorously shown to follow step-by-step from plausible starting assumptions and observation on the local behaviour of light and free particles; as a substantial part of this, the EPS approach establishes the *existence* of a Weyl structure given suitable projective and conformal structures. Compare this to Weyl's original work: the theorem by Weyl—usually referred to as 'Weyl's theorem' in the current context—states only that the Weyl structure is uniquely *determined* by the projective and conformal structure of a presupposed metric; so, Weyl's theorem is an uniqueness result about a Weyl metric, but is not an existence result—let alone an explicit construction of a Weyl metric.[6]

---

[4]In modern-day parlance, this assumption is tantamount to what is known as the *clock hypothesis*: see (Maudlin, 2012, p. 76) for discussion.

[5]This point was made by Weyl himself, see Weyl (1922), p. 63, and so not at all novel to the constructive approaches yet to come.

[6]This existence result follows rigorously only when one supplements EPS with recent results from Matveev and Scholz (2020)—see below.

Over the years, the EPS scheme has found amendments in various directions. Most importantly, proposals for a more admissible basis for the constructive axiomatisation have been put forward (for instance, Woodhouse (1973) and Audretsch and Lämmerzahl (1991)), and problematic steps in the original scheme remedied. With respect to the latter point, in particular, the following achievements are worth highlighting: (1) an arguably highly problematic circularity in determining projective structure was (supposedly) addressed by Coleman and Korté (1992); (2) the apparent reliance of the EPS scheme on primitive clocks in determining the conformal factor was circumvented through a clock construction in Weylian (affine) space by Perlick (1987) (cf. Schmidt (1995)); and (3) the existence of a Weylian metric in the presence of conformal and projective structure was only rigorously proven recently by Matveev and Scholz (2020). An important albeit somewhat irritating result brought up by the EPS scheme consists furthermore in the fact that (4) the details of the constructed spacetime structure depend crucially on certain geometrical assumptions which are hard to justify. For instance, for a given set of local observations, the EPS scheme can lead to some Finslerian-non-Lorentzian spacetime if the twice differentiability assumption of the echo function $p \to t_e \cdot t_r$ ($t_e$ and $t_r$ are emission and return 'times' respectively) is dropped (see Pfeifer (2019); Lämmerzahl and Perlick (2018)).[7]

Our purpose in this article is to provide a comprehensive appraisal of the potential—but also the limitations—of a constructive methodology *qua* epistemology of spacetime theories. For this, we take stock on the plethora of variants on the causal-inertial methods, and related constructive approaches to spacetime theories. We focus on constructive approaches to general relativity, but some of the results pertain to neighbouring or generalised theories of general relativity (e.g. teleparallel gravity) just as well. In fact, our undertaking may be read as a case-study on the merits of a constructive approach to a physical theory more generally.

## 1.3 Plan for the paper

The plan for the paper is this. §2 discusses the 'causal-inertial' constructive axiomatic take on general relativity, with (as discussed above) its roots in Weyl's theorem, and taking its mature form in the EPS axiomatisation. After clarifying the relationships between Weyl's work and the EPS approach, we introduce and assess the (supposedly) circularity-free refinement of EPS developed by Coleman and Korté, as well as Perlick's later refinements to the programme. In §3, we consider the interplay between chronometry (i.e., appeals to rods and clocks) and constructive axiomatics: is it indeed the case that constructive axiomatics such as that of EPS need make no appeal to such complex constructions, or is their use, rather, implicit in the approaches—and if so, where is it implicit? §4 discusses various notions of constructive axiomatisation and constructivism. We consider (i) the extent to which constructivist approaches may legitimately make reference to theoretical notions associated with external theories (i.e. drop the constructive but keep the constructivist nature), and (ii) whether there could be a sense in which constructive axiomatisation can proceed via iterative, rather than linear, methodologies (i.e. drop the constructivist but keep the constructive nature)). The section also treats the question of (iii) how constructive axiomatics relates to the idea of 'Einstein-Feigl completeness': an issue previously discussed by Carrier (1990).[8] §5 situates constructive axiomatics in the broader philosophy of spacetime and of physics, by drawing connections between constructive axiomatics and a number of other programmes and issues. The final section—§6—raises in the context of constructive axiomatics the general epistemological question of how viable a foundationalism of a physical theory can be *per se*.

---

[7]This issue is also discussed in the companion paper, (Linnemann and Read, 2021a).

## 2 Causal-inertial-type constructivism

In this section, we present and discuss three important projects which form part of the history of the EPS scheme. In §2.1, we explain the exact sense in which this approach is related to a series of theorems by Weyl, proved in the 1920s. In §2.2, we consider the attempt by Coleman and Korté to repair an apparent circularity in the EPS approach, and consider whether these authors' revised version of EPS evades that circularity problem. In §2.3, we discuss Perlick's proposed amendments to the EPS programme.[9]

### 2.1 The original motivation: Weyl's theorem

The constructive axiomatic route of EPS is indebted heavily to a result by Weyl, according to which the Weyl metric tensor on a differentiable manifold is uniquely determined by its projective and conformal structure:[10]

> Projective and conformal structure of a [Weyl] metric space uniquely determine the [Weyl] metric (Weyl (1921), Satz 1)[11,12]

---

[8] Roughly for now: a theory is Einstein-Feigl complete just in case it contains its own observation theory.

[9] Another catalogue of the various further modifications which have been made to the EPS approach in the intervening fifty years since its appearance in the literature can be found in (Linnemann and Read, 2021a, §8.2).

[10] Kretschmann (1917) has worked on this as well, and might even have arrived at a similar result prior to Weyl (see Coleman and Schmidt (1995), p. 1343).

[11] The German original reads: "'Projektive und konforme Beschaffenheit eines [Weyl-]metrischen Raums bestimmen dessen [Weyl-]Metrik eindeutig." (Weyl (1921), Satz 1). Although other propositions ('Sätze') are proven in this paper, it is Satz 1 which generally is referred to as 'Weyl's theorem' in this context.

[12] As an immediate corollary it follows for Lorentzian metrics:

**Corollary 1** *Assume $g'_{ab} = \Omega^2 g_{ab}$. Further, assume $g'_{ab}$ and $g_{ab}$ agree as to which smooth, timelike curves can be reparameterized so as to be geodesics. Then $\Omega$ is constant. (Malament (2012), proposition 2.1.4, p. 127)*

The result guarantees that if suitable projective structure and conformal structure that give rise together to a Weyl metric are found, *then* the Weyl metric would be uniquely determined by these. Notably, the result is not an existence result for a Weyl metric in the presence of conformal and projective structure.

Now, the causal-inertial method—as the name already suggests—aims to construct from causal structure (in the sense of conformal structure) and inertial structure (in the sense of projective structure) a (Weyl) metric structure. In the EPS approach, just as the Weyl metric is seen as uniquely determined by its projective and conformal structure (Weyl's theorem), projective and conformal structure are seen as determined by their respective geodesic structure. It is important to stress the distinction of the latter existence result from the former uniqueness result (which is behind Weyl's theorem); the existence result is a core result of the EPS scheme: provided that there is projective and conformal structure such that the null lines of the conformal structure are projective geodesics (autoparallels) of the projective structure, both structures together define a Weyl structure.[13]

Now, in relation to EPS, it is interesting to note that conformal structure in particular is determined uniquely by (at least) one other criterion than the corresponding structure of null geodesic paths—namely, by the path structure of all timelike curves (Malament, 1977).[14] How do the two options compare operationally? (Let us assume that the only operationally sensible method to measure out events from one specific location is via radar coordinates, and that the only reliable standard of constant signal speed in curved spacetime is that of light.[16]) One advantage of this new 'timelike curve' option over EPS' original approach lies in the fact that it is operationally easier to measure out the paths of timelike curves in such radar coordinates than the paths of lightlike signals which themselves move at the speed of light. The disadvantage of this new approach, on the other hand, is that many more curves have to be probed to obtain a good resolution of the conformal structure (figuratively speaking, the inside of the light cone has to be measured out as opposed to just its boundary).[17]

---

[13] The existence result has been been given a rigorous formulation by Matveev and Scholz (2020).

[14] In total, and more formally, two metric fields $g'_{ab}$ and $g_{ab}$ are conformally equivalent iff

**Lightcone structure**  $g'_{ab}$ and $g_{ab}$ agree on which vectors, at arbitrary points of $M$, are timelike (or agree on which are null, or which are causal, or which are spacelike).

**Timelike path structure**  $g'_{ab}$ and $g_{ab}$ agree on which smooth curves on $M$ are timelike.[15]

**Null geodesic structure**  $g'_{ab}$ and $g_{ab}$ agree on which smooth curves on $M$ can be reparameterized so as to be null geodesics.

(This is a combination of the proposition 2.1.1. (p. 122), and proposition 2.1.2. (p. 125) from Malament (2012)).

[16] This isn't necessarily completely uncontroversial, at least from the point of view of the completed theory of general relativity—see e.g. Asenjo and Hojman (2017); Linnemann and Read (2021b).

[17] Here is an interesting technical question: given a finite subset of all continuous timelike curves,

## 2.2 Circularity-free EPS: Coleman and Korté's account

In the wake of EPS' paper, the charge was quickly raised that the approach suffers from a fatal circularity, insofar as it requires substantial conventionalist *inputs* regarding spacetime geometry. Here is how Sklar put the point:

> ... is a particle free or not when it is gravitationally attracted by another particle? If we say it is, and use the EPS construction, we shall get one spacetime. But what if we say it is not? Won't we get a whole "conventionally alternative" space-time by using the EPS construction? (Sklar, 1977, p. 259)

Essentially, the issue is this: how is one to identify the freely-falling particles—an input in the EPS construction—without an antecedent understanding of spatiotemporal structure used to underwrite this notion? Motivated by this concern, Coleman and Korté developed, over the course of a series of papers, a (supposedly) non-circular modification of the EPS axiomatisation. In this section, we provide a summary of the Coleman-Korté 'improvement' of EPS, as presented in Coleman and Korté (1992).[18]

Letting the acceleration of a particular point particle ('monopole') be $\xi^b \nabla_b \xi^a$, where $\xi^a$ is the velocity vector of the particle under consideration, and $\nabla$ is a derivative operator on the differentiable manifold, Coleman and Korté introduce a *directing field* $\Xi^a$, as that term on the right-hand side of Newton's second law:

$$\xi^b \nabla_b \xi^a = \Xi^a. \tag{1}$$

How, though, is the directing field to be identified operationally? In order to answer this question, Coleman and Korté introduce what they dub the *monopole criterion* (Coleman and Korté, 1995b, p. 176):

> If two monopoles of the same class are launched at nearly the same event with nearly the same 3-velocity, then under all external physical conditions, their (future) worldline paths will be nearly the same.

By "monopoles of the same class" is meant monopoles subject to the same directing field on the right-hand side of (1). The monopole criterion, then, states that if two particles (monopoles) are assigned the same directing field, then they must follow approximately the same trajectories, for approximately the same initial conditions. Even more simply: particles subject to the same forces and with roughly the same initial conditions should behave in roughly the same way.[19] Coleman and Korté present a procedure via which, given the monopole criterion, a directing field can be *constructed* (Coleman and Korté,

---

how accurately can a given piece of conformal structure be recovered?

[18]See also the excellent summary by Coleman and Schmidt (1995).

[19]Note that, at least in (Coleman and Korté, 1995a,b), Coleman and Korté do not assert the reverse implication: that particles which behave in the same way should be subject to the same forces. This will be of relevance below. (See footnote 21.)

1995b, pp. 177-178). In this way, as presented above, this is claimed to afford a non-circular means of accessing inertial structure (what Coleman and Korté dub the 'cubic criterion' is used to select a privileged monopole class).

Even if one avers that the Coleman-Korté approach succeeds in this regard, we contend that one should be wary of any claims to the effect that this renders spacetime geometry non-conventional. One reason has already been articulated by Pitts, who writes that

> Unfortunately the anti-conventionalist argument hinges entirely on the assumption that there is exactly one physically relevant conformal metric density and exactly one physically relevant projective connection. (Pitts, 2016, p. 84)

This is exactly correct. The point here is: it is possible that the Coleman-Korté project leads one to construct physical theories with *multiple* spacetime geometries. In that case, which is the "One True Geometry" (Pitts, 2019, p. 5) is underdetermined, and, in this particular regard, issues of conventionality re-arise. (More on this below.)

There is also a second possible source of conventionality in the Coleman-Korté approach. As these authors acknowledge, "It is important to emphasize that all of the particles used in the measurement procedure must be known to belong to a specific monopole class *before* they are used in the measurement procedure" (Coleman and Korté, 1995b, fn. 7). But this means that there might, in principle, be multiple different possible partitionings of the subject class of monopoles into sub-classes with the same directing fields. We might, in particular, identify certain *different* classes of monopoles as forced versus force-free, leading to the identification of differential inertial structure, and so (potentially) different spacetime structure. The reasons for this are, essentially, twofold. First: how is one to determine whether two monopoles, spatially *separated* and with *different* initial velocities, should belong to the same monopole class or not? According to the monopole criterion, they may be regarded as belonging to the same such class if it is the case that, *were* their initial conditions and velocities roughly the same, they *would* behave in approximately the same manner. But, clearly, to state whether this is the case or not is to assert certain modal facts which transcend the empirically accessible data—it is, ultimately, to make an empirically ungrounded *assertion* about inertial structure. Given the possibility of giving different answers to this modal question, it should be clear that the Coleman-Korté prescription does not evade all issues of conventionality.[20] Second: Coleman and Korté assert that monopoles in the same class

---

[20]The issue here is that the Coleman-Korté operationalisation is *local* and *in principle*: given, locally, the motions of a sufficiently large (potentially practically unrealisable) number of test bodies, one may or may not be able to fix uniquely a notion of forced versus unforced motion (this is what Coleman and Korté mean when they describe the law of inertia as an "empirical law" (Coleman and Korté, 1982, p. 268)): interestingly, *if* spacetime structure is Lorentzan, such is the case; not so if spacetime structure is Newtonian (see Coleman and Kort (1995)). However, *this is all local*: if one is, from one's 'God's eye view', attempting to use the motions of *all* bodies to reconstruct inertial structure *in toto*, Reichenbachian 'coordinating definitions' (see (Reichenbach, 1969, ch. 1)) must be

and with approximately the same initial conditions should behave in approximately the same manner. They do not, however, assert that monopoles with approximately the same initial conditions which behave in approximately the same manner must be assigned to the same monopole classes—i.e., they do not assert that such monopoles must be regarded as being subject to the same directing field. In this case, again, there arises conventional choice as to whether to regard monopoles as belonging to the same class, and so there arises conventional choice about the fields and objects which, ultimately, correspond to inertial structure.[21] Overall, then, while their strategy is ingenious and illuminating, insofar as Coleman and Korté claim to evade *all* issues of conventionality, they are not correct.

It is also worth exploring the differences between this conventionality in one's choice of directing field with conventionality of the Pittsian kind. In a sense, the former is closer to the original Reichenbachian concerns regarding the conventionality of geometry: the issue is that there are different possible ascriptions of inertial structure (and so geometrical and spacetime structure) to the world, based upon certain initial conventional choices. Conventionality of the Pittsian kind is, rather, principally an issue of nomenclature: there is one (perhaps rather complex, e.g. multi-metric) geometrical structure to the world, but within that structure, *which* object is to be dubbed 'spacetime' is underdetermined.[22] Only the former of these cases of conventionality presents a substantive metaphysical concern about the structure and ontology of the world.

The Coleman-Korté response to Sklar's conventionality challenge to EPS is certainly illuminating; however, in light of the above, the jury is out as to whether it is fully successful. We should also flag they there is a range of less mathematically heavy-duty approaches to identifying inertial trajectories (i.e., to identifying the freely-falling bodies) without presupposing spatiotemporal structure: see (Read, 2022, ch. 1) for a recent review.

At this point, it is worth mentioning a related (apparent) programmatic circularity in setting up selection criteria for geodesics (as proposed by Coleman-Korté), identified by Carrier:

> the theory in fact enters the scheme even if only in a hidden or indirect fashion. Why, after all, do we regard trajectories influenced by gravitational quadrupole moments as distorted whereas we consider paths determined by the action of gravitational monopoles as perturbation-free? Evidently, it is the explanatory theory (GTR) that

---

introduced *vis-à-vis* whether distant particles are regarded as being subject to the same directing field, and so the spectre of conventionality re-emerges.

[21]Unlike Coleman and Korté (1995a,b), at (Coleman and Korté, 1980, p. 1350) (an older piece) this condition *is* asserted as a biconditional—in which case, the second of these two concerns does not apply. A charitable reading here would be to take Coleman and Korté to be deploying the 'mathematician's if'—in which case, this concern should be set aside.

[22]In order to answer this question, one might appeal to e.g. the 'spacetime functionalism' of Knox (2019). Of course, inertia still comes in here: the question is: which of the two (or more) pieces of geometrical structure picks out the 'true' inertial motions?

induces and justifies this judgment. Theory enters constructive axiomatization by *defining the ideal cases*. It supplies *criteria of adequacy* for our choice of basic processes. Only because we already know GTR we can feign to be ignorant of it. (Carrier, 1990, p. 380)

Carrier's point is that one only adheres to a correction scheme such as that of Coleman-Korté if one already has reason for preferring certain idealised free-fall structure—as opposed to some structure distorted relative to it—as decisive.[23,24] To stress: Carrier's complaint is not that, once an improvement scheme *à la* Coleman-Korté has been chosen, it is itself in its application circular; rather, the circularity which he identifies exists (supposedly) at the programmatic level. Now, in fact, it is not at all clear that a programmatic dependence on some desired target theory—as arguably the case here—truly deserves the label 'circularity' (of whatever form): after all, it is not the case that a constructive axiomatic endeavour leads to a specific (family of) theories just because *some* structure is construed with *some* theory goal in mind at *some* intermediate step of the overall theory construction. In other words, the programmatic circularity attested to the improved EPS-scheme seems, in our view, to be non-vicious (albeit not fully in the desired constructivist's spirit).

## 2.3 The local constraints of the radar method

The radar method is only available in a neighbourhood around a point. How practical, then, is the radar method? Before we attempt to answer this question, note that there are basically two distinct radar methods referred to in the context of EPS: (i) that of EPS, and (ii) that in the spirit of Perlick (1987). While that of EPS requires two observers, that of Perlick is a straightforward one-observer procedure: radar method (i) fixes points through the intersection of emitted and received signals of one observer with that of another, whereas radar method (ii) fixes points through the intersection of an emitted signal (that is, implicitly, taken to be singled out further through two directional angles) with the reflected signal from those points alone.

One might simply claim that Perlick's scheme is thereby more tractable operationally than the original EPS radar method: the problem with method (i), after all, is that it requires communication of the two observer systems: each observer must send (with some appropriate encoding) to the other observer her

---

[23](Carrier, 1990, pp. 379–380) distinguishes two kinds of correction: by a theory different from that considered specifically (e.g. electrodynamics), and by the theory considered itself.

[24]Cf. Hilbert's remarks on the purpose of axiomatising physical theories:

> The edifice of science is not raised like a dwelling, in which the foundations are first firmly laid and only then one proceeds to construct and to enlarge the rooms. Science prefers to secure as soon as possible comfortable spaces to wander around and only subsequently, when signs appear here and there that the loose foundations are not able to sustain the expansion of the rooms, it sets about supporting and fortifying them. This is not a weakness, but rather the right and healthy path of development. (Corry, 2004, p. 127)

For further fascinating discussion of Hilbert on these issues, see Corry (2018).

two measurement values (i.e., emission and reception times) so that the other observer can complete her respective set of coordinates for the event in question (which consists of the respective emission and reception times on both observers' worldlines for the event in question). Notably, though, an intra-observer communication becomes necessary even on Perlick's account of radar coordinates (method (ii)) as soon as the EPS scheme is meant to reconstruct more than just the local kinematical arena of one observer.

How, for instance, to relate two radar coordinate maps of the Perlick type associated to a pair of neighbouring, arbitrarily-parameterised wordlines in some generic spacetime? A straightforward procedure might run as follows:

1. Send a light ray $L_1^E$ from worldline $\gamma_1$ towards wordline $\gamma_2$; let the light ray hereby encode the emission parameter time $t_1^E$.[25]

2. Record the parameter time of reception for $L_1^E$ on $\gamma_2$ (denoted by $t_2^R$).

3. Send back a light ray from $\gamma_2$ as the reflection of $L_1^E$ (denoted by $L_1^R$); let this light ray $L_1^R$ hereby encode the values of $t_1^E$ as well as of $t_2^R$.

4. Record the parameter time of reception at $\gamma_1$ (denoted by $t_1^R$).

5. The observer at $\gamma_1$ can now associate an event on $\gamma_2$ with two of her own parameter values for emission and receival ($t_1^E$, $t_1^R$) as well as the parameter value of that event for $\gamma_2$ ($t_2^R$).

(Especially noticeable in non-static spacetimes—i.e., spacetimes which are not approximately static on the relevant time-scales—is the fact that one would have to constantly keep tiling spacetime with such a light signal exchange in order to guarantee the interchangeability between neighbouring radar coordinates, which is, admittedly, far from practical.) Again, it bears stressing that the encoding proposed here represents yet another quite explicit instance of *external* theory-dependence.[26]

## 3 Chronometry and constructivism?

Chronometric approaches to GR treat rods and clocks as primitive objects, whereas the proponents of constructive axiomatisations like EPS typically claim that their methodology does not require us to take complex objects of this kind for granted (see e.g. the introduction to Ehlers et al. (2012)). But it has been argued that EPS and other constructive axiomatisations do in fact appeal implicitly to clocks—in particular, in the course of moving from a Weyl metric to a

---

[25]Some external theory will describe the encoding—so theory-neutrality is also lost here. Note that this is at least in line with Reichenbach's more permissive idea of *constructive* axiomatics in which the constructive vocabulary may be subject to theoretical terms as long as the employed terms are different from the theory to be constructed.

[26]As we have already seen in the above footnote, though, in a constructivist account *à la* Reichenbach (1969) this is unproblematic.

Lorentzian metric EPS use a 'no second-clock' criterion which arguably presupposes the availability of clocks as primitive objects. In this section, we consider various approaches to making sense of rods and clocks in a constructive axiomatisation, and assess the extent to which they can be employed within the EPS scheme in order to obviate the need to rely on primitive rods and clocks.

### 3.1  *Bona fide* chronometry

It is probably correct to endow Synge with the title of the most dedicated proponent of a (primitive) chronometric take on GR: in his seminal textbook (1960) he puts a primitive correspondence between (the general relativistic notion of) proper time and the reading of some sufficiently good theory-external clock (such as an atomic clock)—dubbed a 'standard clock'—as center-stage for any empirical account of the metric field; length measurements are then derivative on time measurements under the assumption of a constant speed of light.[27] Now, it is with accepting clocks as theory-external objects in the interpretation of GR that EPS (and their followers) take issue. Notably though, there is another, in fact distinct, tradition of criticising presupposed theoretical clocks in GR: its progenitor Einstein, and, among others, Feigl more generally (see, for instance, (Feigl, 1950)), demanded that a theory provide a description of its 'observation theory'—including e.g. its clock constructions—without, however, any demand for constructivism or constructive axiomatisation more specifically. Carrier summarises succinctly the distinction between the two traditions as follows:

> The constitutive principles of constructive axiomatizations are (1) the methodological requirement of direct testability and (2) the semantical postulate of linearity. ... EFC [Einstein-Feigl completeness], on the other hand, is guided (1) by the methodological idea of explanatory power (that is, the idea that a theory should explain a large range of phenomena in a precise fashion by invoking as few independent assumptions as possible) and (2) by a semantical account that allows for reciprocal clarification of concepts and theories. ... Concerning the drawbacks of both approaches, note that the whole observation basis of a theory cannot be constructively axiomatized and that complete theories (if only in extreme cases) may suffer from test restrictions (Carrier, 1990, p. 391)[28]

As noted in the introduction, *bona fide* chronometry *qua* epistemology of spacetime is *prima facie* less ambitious than the causal-inertial method insofar as the import of an external, non-primitive element to GR ruins any prospects of an Einstein-Feigl completeness—the idea that the theory provides its own observational theory—from the start.[29, 30] Note, therefore, that chronometry—and, indeed, any theory which is not Einstein-Feigl complete—is susceptible to the charge of 'legitimising sin', in Einstein's sense (again, see Giovanelli (2014) for further discussion).[31]

---

[27] Again, it's not obvious how uncontroversial this assumption really is: see Asenjo and Hojman (2017); Menon et al. (2018); Linnemann and Read (2021b).

[28] In the same passage, Carrier emphasises nicely the methodological advantages of the Einstein-Feigl tradition over that of EPS (and constructive axiomatisation more generally).

## 3.2 Weylian (inertial) chronometry

The EPS scheme uses a clock criterion ('no second clock effect') to reduce a Weyl spacetime to a Lorentzian spacetime in the axiomatic process. We will now critically review clock accounts for a Weyl spacetime context with regard to their applicability within the EPS scheme: such accounts have been developed with a view to filling this lacuna in the original EPS article.[32] We will structure the discussion of clocks in the context of Weyl spacetimes along two traditions, namely one of operationalist constructions and one of operationalist test criteria of clocks. Before we can discuss such clock accounts, we have to clarify what it means to have a notion of time admissible to clock measurements in a Weylian context to begin with—without a relevant notion of time, after all, there is nothing for clocks to measure.

### 3.2.1 Proper time in Weyl spacetime

In this subsection, we introduce the (not very well-known) notion of proper time for Weyl spacetimes. A notion of proper time for Weyl spacetime cannot be defined in direct analogy to the worldline interval in Lorentzian spacetimes, as such an object would not be invariant under Weyl (i.e., spacetime-dependent scale) transformations. Following Perlick (1987); Avalos et al. (2018), a suitable notion of proper time in Weyl spacetime that is extensionally equivalent to the standard notion when restricted to Lorentzian spacetimes can, however, be set up as follows:

1. Consider the distinction between geodesics proper and geodesics in the sense of pregeodesics. A pregeodesic differs from a proper geodesic insofar as the proportionality constant between the tangent vector and its change relative to itself can be non-zero, i.e., $\frac{D\gamma'(u)}{du} = f(\gamma(u))\gamma'(u)$ with $f(\gamma(u))$ arbitrary for a pregeodesic, and $f(\gamma(u)) = 0$ for a geodesic. (One may now simply define a notion of proper time for geodesics in Weyl spacetime as that parametrisation of a pregeodesic that makes it a geodesic—but this notion cannot be directly taken over to arbitrary worldlines.)

---

[29] It has been said before that chronometry picks out time as preferred for epistemic purposes over space. Note that proper time is a preferred notion in general relativity from an observer's point of view over that of proper length: the former is arbitrarily defined along her whole worldline whereas the latter is just a reasonable concept within a small spacelike neighbourhood intersecting her worldine alone.

[30] The chronometric approach is one fall-back option in all those spacetimes which cannot come out of a constructivist axiomatisation, say when no light rays nor any other sense of field moving approximately close to the null cone are available. Note though that this does not entail yet that tracking is impossible in such spacetimes without the chronometric approach: the Einstein simultaneity condition via light signals is not the only sensible operationalist manner to arrive at an operationalist coordinate system for tracking. See for instance the supplements of Janis (2018) and references therein for how to set up a standard of synchrony from clock transport alone.

[31] Note that Einstein-Feigl completeness accords which Stein's injunction to 'schematise the observer' in physical theories: see (Stein, 1994), and (Curiel, 2019) for further discussion.

[32] For further discussion of this issue, see (Linnemann and Read, 2021a).

2. Importantly, the geodesic criterion for proper time $\frac{D\gamma'(u)}{du} = 0$ is equivalent to $g(\gamma'(\tau), \frac{D\gamma'(\tau)}{d\tau}) = 0$ where $g$ is a representative of the Weyl metric. This orthogonality condition, can, in contrast to the proper time criterion based on the (pre)geodesic equation, be applied to timelike curves as well.

3. Thus, define a timelike curve in Weyl spacetime as parametrised by proper time $u$ for a segment between two events iff $\frac{D\gamma'}{du}$ is orthogonal to $\gamma'(u)$ for all $u$ in that segment. (As stressed before, it can be shown explicitly that this notion of Weylian proper time reduces to the standard notions of proper time for Lorentzian and Weyl-integrable spacetimes more generally.)

Compare this to the definition of proper time for Weyl spacetime taken up by EPS, according to which "a time-like curve $\gamma$ is parameterized by proper time if the tangent vector $\gamma'$ is congruent at each point of the curve to a non-null vector $V$ which is parallel-transported along $\gamma$" (Avalos et al., 2018, p. 26), i.e. $g_p(\gamma'_p, \gamma'_p) = g_p(V_p, V_p)$. While it can be proved that this definition is equivalent to that of Perlick (see Avalos et al. (2018), proposition 2), Avalos et al. (2018) have a point in stressing that Perlick's presentation is more satisfactory, as it is a definition of proper time explicitly in terms of the worldline's properties alone—rather than making adherence to rather abstract 'comparison' vector fields.

We will now consider the two traditions of clock considerations—one which aims at explicit clock constructions, and another which provides a test criterion for clocks.

### 3.2.2 The Langevin clock construction

Put simply, the idea is to construct a clock from light rays or particles—i.e. types of idealised motion without back-reaction—that move back and forth in a well-defined segment and thereby reads out a suitable (generalised) notion of proper time. For Lorentzian spacetimes, arguably the most rigorous existing treatment of Langevin clocks is that of Fletcher (2013), who provides a theorem on how light rays in a light clock moving on a time-like curve measure out some quantity proportional to proper time between two ticks (accuracy requirement), and do so in a regular fashion, i.e., the interval between two ticks is always the same (regularity requirement).[33] Correspondingly, for Weyl space-

---

[33] The tradition of rigorous light clock constructions within general relativity at least goes back to Marzke and Wheeler (1964) who construct a light clock for Lorentzian spacetimes, with the

time, Köhler (1978) has shown that light clocks moving along pregeodesics fulfil the regularity requirement.[34]

The problem with all these constructions (and ultimately the reason why they cannot help in the EPS approach) is that they presuppose a notion of spatial regularity in order to latch onto the wordline they are supposed to measure out. In the Lorentzian case, this means that one needs to presuppose what is known as Born-rigidity to hold between both sides of the clock. In the Weylian case, the problem is even more acute: first, there is no naïve sense of constant length available due to Weyl scaling. Second, note that the Weylian notion of proper time is non-geometric, i.e., it is not just an integral over local features but an one over features that each depend on the starting position of the clock; after all, the explicit formula for the proper time between the parameter values $t$ and $t_0$ of a curve in Weyl spacetime as discussed above is given by (Avalos et al., 2018, p. 259):

$$\Delta\tau(t) = \frac{\frac{d\tau(t_0)}{dt}}{(-g(\gamma'(t_0), \gamma'(t_0)))^{\frac{1}{2}}} \int_{t_0}^{t} e^{-1/2 \int_{t_0}^{u} \omega(\gamma'(s))ds} (-g(\gamma'(u), \gamma'(u)))^{\frac{1}{2}} du. \quad (2)$$

A notion of proper *length* for Weyl spacetimes—obtainable via the presumption of constant speed of light—is thus presumably non-geometric (in the sense just defined above) as well.

### 3.2.3 Clock criterion

We saw that a clock construction that is to show temporal regularity presupposes spatial regularity. A way out is then to relax the constructivist requirement and simply demand a criterion for distinguishing a good clock from a bad one. In fact, a clock criterion might under some circumstances still be seen a part of a constructivist approach—albeit not of a linear but rather of an iterative constructivist approach (more on which below): start with some arbitrary Langevin setup, then tweak this 'clock' in some way: if the clock works better than before according to the criterion, then continue with that clock and tweak it again (cf. Tal (2016)). However, not only is convergence to a standard clock not guaranteed, but even if the iterative procedure does converge, importantly, the EPS scheme will not be restored this way as a linear constructivist project.

In this subsection, we introduce Perlick's construction which, using only facts about light propagation, allows us to test an arbitrary parameterisation of a timelike curve to determine whether it can be associated with a standard clock. Consider a general relativistic spacetime $(M, g_{ab})$ with a time orientation—i.e., a globally-consistent distinction between future and past (see (Malament, 2012, ch. 2)). Following Perlick (Perlick, 1987, p. 1), define a *clock* as follows:

---

restriction that the light clocks move on geodesics; see also Desloge (1989). For some philosophical discussion of Fletcher's clock construction, see (Menon et al., 2018).

[34]An alternative result for Langevin-type clock constructions in Weyl spacetimes is by Castagnino (1968) who for any torsion-free affine geometry (which include any Weyl spacetimes) provides clock constructions.

**Definition 3.1 (Clock)** *A smooth embedding $\gamma : t \mapsto \gamma(t)$ from a real interval into $M$ such that the tangent vector $\dot{\gamma}(t)$ is everywhere timelike with respect to $g_{ab}$ and future-pointing.*

Perlick is correct in stressing that the above definition captures the essential theoretical aspects of a clock—for any timelike curve may in principle represent the worldline of a clock; in particular, one may interpret the value of the parameter $t$ as the reading of that clock. As Perlick also stresses, nothing in this definition requires that the clock be a *good* clock—in the sense that nothing requires the ticking of the clock (i.e., the choice of parameterisation) to be 'uniform', or to coincide with the regularities exhibited by material systems.

Such a clock in hand, Perlick then uses the 'radar method' to 'spread time through space'—i.e., to assign coordinates *off* the worldline of the clock, at some point $q \in M$. This proceeds as follows (recall also our above discussion of the radar method *chez* Perlick). First, consider the curve $\gamma$ representing this clock. One emits a light ray from an event on $\gamma$, say $\gamma(t_1)$; this light ray is reflected from $q$ and is received back at some later point on $\gamma$, say $\gamma(t_2)$. (Note that this method *presupposes* that there is some means of reflecting the signal at $q$.) Then, the *radar time $T$* and *radar distance $R$* of $q$ from $\gamma$ are defined as follows:

$$T = \frac{1}{2}(t_2 + t_1), \tag{3}$$

$$R = \frac{1}{2}(t_2 - t_1). \tag{4}$$

Three assumptions here are worth stressing. First: we assume that the dynamics of the light signal isn't affected by the geometry of the spacetime under consideration—this, of course, is a non-trivial assumption.[35] Second: this method of setting up radar coordinates presupposes the Einstein-Poincaré clock synchrony convention (see e.g. (Reichenbach, 1956) for discussion), insofar as $q$ is assigned a temporal coordinate *half-way between* $t_1$ and $t_2$ on $\gamma$. This latter assumption is acknowledged explicitly by Perlick, when he writes that

> Incidentally, if one replaces [(3)] by $T = pt_1 + (1-p)t_2$ with any number $p$ between $0$ and $1$, each hypersurface $T$ = constant gets a conic singularity at the intersection point with $\gamma$. This clearly shows that the choice of the factor $1/2$ is the most natural and the most convenient one. (If one allows for direction-dependent factors, one can get smooth hypersurfaces with factors other than $1/2$. This idea, which however seems a little bit contrived, was worked out by Havas [Havas (1987)] where the reader can find more on the "conventionalism debate" around the factor $1/2$.)

To be clear: the above more general choice for $T$, with direction-independent factors other than $1/2$, goes back to Reichenbach, and is now known as the

---

[35]In light of this asumption, one should treat with suspicion Perlick's writing that "Adopting the standard formalism of general relativity, "light ray" is then just another word for "lightlike geodesic of the spacetime metric $g$"." (Perlick, 1987, p. 2)

*Reichenbach-I synchrony convention.* On the other hand, a choice of direction-independent factors *à la* Havas is now known as the *Reichenbach-II synchrony convention.*[36]

The third assumption made above is that there exists a mirror at $q$ capable of effecting the reflection—clearly, this is a substantive assumption, which will mandate prior coordination on the part of the experimentalist. (Note also that a mirror is not strictly necessary to set up these radar coordinates, and that any other method of ensuring that the signal return from $q$ to $\gamma$ would suffice—e.g., a fly-by scenario where the incoming signal is crossed by another signal which effectively looks like a reflection of the incoming signal at a mirror.)[37]

As Perlick stresses, although the radar method fails globally in general relativistic spacetimes, it always works locally. This fact is captured in the following proposition (Perlick, 1987, p. 4):[38]

**Proposition 1** *Let $\gamma$ be a clock in an arbitrary general relativistic spacetime, and let $p = \gamma(t_0)$ be some point on $\gamma$. Then there are open subsets $U$ and $V$ of the spacetime with $p \in U \subset V$ such that every point $q$ in $U \setminus Im(\gamma)$ can be connected to the worldline of $\gamma$ by precisely one future-pointing and precisely one past-pointing light ray that stays within $V$. In this case, $U$ is called a* radar neighbourhood *of $p$ with respect to $\gamma$.*

We now introduce Perlick's notion of a *standard clock*:

**Definition 3.2 (Standard clock)** *Let $\gamma$ be a clock, and let $\xi^a$ be the tangent vector field to $\gamma$. Then a clock is a* standard clock *with respect to a given metric field $g_{ab}$ just in case $g_{ab}\xi^a\xi^b = -1$. Then the parameter of the clock is called the* proper time.

Note that standard clocks 'march in step' with the metric field—i.e., the satisfy the clock hypothesis (already discussed above).

Now, Perlick claims that the radar method affords a means of ascertaining whether a given clock is a standard clock (with respect to some metric field). The procedure works as follows. Consider a clock $\gamma$, and two freely-falling particles $\mu$ and $\bar{\mu}$ emitted from $\gamma$. The radar method assigns a time $T$ and distance $R$ to each event on $\mu$, and a time $\bar{T}$ and distance $\bar{R}$ to each event on $\bar{\mu}$. Now, the standard clock condition holds at $t = t_0$ (which corresponds to $T = \bar{T} = t_0$ just in case

$$\frac{\frac{d^2R}{dT^2}}{1 - \left(\frac{dR}{dT}\right)^2}\Big|_{T=t_0} = -\frac{\frac{d^2\bar{R}}{d\bar{T}^2}}{1 - \left(\frac{d\bar{R}}{d\bar{T}}\right)^2}\Big|_{\bar{T}=t_0}. \tag{5}$$

---

[36]See (Read, 2022, ch. 7) for further discussion, including of how conic singularities can be avoided in the latter case.

[37]One would not need to make this assumption if one interpreted the radar coordinates in a modal sense—i.e., if there *were* a mirror at $q$ then it *would* be reflected to $\gamma$. For more on such 'modal' constructivist approaches, see Adlam et al. (2022); for the time being, we leave it open whether to appeal to modal notions is truly in the constructivist spirit. Note also that EPS themselves make liberal use of coordinates in their article—so they, too, seem to be invoking some degree of modal understanding of coordinates.

[38]As Perlick himself notes, the existence of radar neighbourhoods in the sense of the proposition below was also assumed in EPS approach. This is discussed in Linnemann and Read (2021a).

Suppose, then, that one experimentally ascertains that (5) obtains. The upshot is that there exists some Lorentzian metric field with respect to which the clock in question is a standard clock. To stress again: this is a *test* of whether a clock is 'standard', rather than an explicit clock *construction* (which was the topic of the previous subsection).

Notably, Hobson and Lasenby (2020, 2022) have recently called into doubt whether the generalised proper time for Weyl spacetime *à la* Perlick is physical. Specifically, they criticise the claim that the orthogonality condition presupposed in the derivation of the proper time formula, $g_{\mu\nu}u^\mu a^\nu = 0$, is required to transform non-physically with $u^\mu \to u^\mu$ under a Weyl transformation, when its actual transformation should be of form $u^\mu \to e^{-f}u^\mu$, where $e^f = \Omega^2$. The standard of physicality invoked by Hobson and Lasenby is, however, debatable:[39] they consider observed quantities necessarily to be described locally by a tetrad system; the measured (and thus physical) velocity measured by the observer is, for instance, supposed to be $u^a = e^a_\mu u^\mu$ (where $a$ denotes the lab index and $\mu$ the usual spacetime index), and not $u^\mu$; given that it is $u^a$ that is taken to be physically relevant, it is also this velocity that has to be invariant under Weyl transformation. With $e^a_\mu$ transforming as $e^f e^a_\mu$, it would indeed follow that $u^\mu$ transforms with $\exp f$, i.e., has Weyl weight $-1$.

Interestingly, such an ascription makes the length of the tangent vector Weyl invariant ($g_{\mu\nu}u^\mu u^\nu$ with $g_{\mu\nu}$ of Weyl weight +2 has a zero total Weyl weight). But—and hence the problem for Perlick's criterion—this means that the notion of proper time *à la* Perlick is no longer physically sensible, as it is no longer invariant under Weyl transformation (if $u^\mu := dx^\mu/d\xi$ has weight $-1$, so does $d/d\xi$ for arbitrary $\xi$—$x^\mu$ after all does not change under Weyl transformation). More than that: (1) Massive particles are—since otherwise not in line with the scale-invariant background geometry—only instantiable through a dynamically coupling auxiliary field. (This requires making recourse to a Lagrangian structure for the particle and thus in a sense to at least placeholder relations to the dynamics already.) (2) The adequate connection is not the standard one as considered by EPS and the related literature but the fully covariant connection $\nabla*_\mu = \nabla_\mu + \omega B_\mu$ where $\omega$ denotes the Weyl weight of the object to act on, and $B_\mu$ is the Weyl potential. Interestingly, though, neither velocity nor acceleration of massive particles that are in free fall are generally parallel propagated relative to the fully covariant connection $\nabla*$.

It is arguably not immediate though that $u^a$ is more physically relevant than $u^\mu$ and should thus count as invariant under Weyl transformations; after all, the tangent vector structure of the manifold—and not the local Minkowski structure associated to the tetrad formalism—may be seen as immediately physically relevant (after all the metric and not the Einstein-Cartan formalism of GR is the default physical rendering of general relativity).[40] But as already pointed out, a mere clock criterion (such as that of Perlick) fails to offer a genuine con-structivist link to Lorentzian spacetimes in any case. The overall leap from Weylian to integrable Weyl/Lorentzian structure thus remains unsatisfactorily resolved when judged from the original constructivist spirit of EPS. Efforts towards more constructivist alternatives to Perlick's clock criterion have either been made at the cost of the constructive nature of the scheme, i.e., the empirical immediateness of the axioms, by accepting (quantum-originating) matter waves, for instance (see Audretsch and Lämmerzahl (1991)), or through switching to an alternative (albeit related) scheme than EPS altogether, which would, however, have to be scrutinised on its own merits (see Schelb (1996c)). A criterion which requires the metric to show a certain symmetry, as already provided by Schelb (1996a), might give rise to the hope that a criterion to differentiate integrable Weyl spacetimes from non-integrable spacetimes in terms of nothing other than radar coordinate events is possible (after all, metric entries do get described as measurable in the EPS scheme), but so far this approach has not been developed satisfactorily.

---

[39] Hobson and Laseby seem to implicity presuppose a form of equivalence principle that would, however, require further discussion.

[40] Arguably, though, this is a weak rebuttal given that some matter (in particular, spinorial matter) requires tetrads for its formulation.

### 3.3 Parameter clocks

The EPS axiomatisation still builds on clocks in a very limited sense, namely on what one might call 'parameter clocks' (as also noted in (Schelb, 1996c, p. 1324)): some arbitrary parametrisation of a wordline might be thought of as serving as a clock in the sense that any monotonously growing function of an object (like the height of a still-growing tree) can do so, even when there is no straightforward notion of periodicity associated with that object.

It should be clear that no objection against the clock-free approach can be mounted from the use of parameter clocks; their assumption commits one to nothing but a time-forward moving process—in other words, the assumption that there is something like time for clocks to be measured to begin with.

To summarise this section, then: there is a (by now well-known) hole in the original EPS axiomatisation, in the sense that EPS assume no second clock effect in order to move from Weyl to Lorentzian spacetimes, but do not provide explicit clock constructions in Weyl spacetimes consistent with the axioms in order to underwrite this claim. Though efforts have been made to make good on such clock constructions in Weyl spacetimes, there remains work to be done on this front; moreover, Perlick's clock *criterion* clearly does not fully realise the constructivist's ambitions in this regard. Finally, EPS also make use in their construction of what we have called parameter clocks, but this should be regarded as being comparatively unproblematic.

## 4 Varieties of constructivism

We now ascend to a more general level. The purpose of this section is to characterise, at the greatest possible level of generality, the different varieties of constructivism. To this end, it is helpful to recall again Carrier's presentation of constructive axiomatics:

> The constitutive principles of constructive axiomatizations are (1) the methodological requirement of direct testability and (2) the semantical postulate of linearity. A theory is to be founded on propositions that are immediately amenable to experience; and the concepts employed by that theory should be clarified first (by relating them to experiences) and afterwards used to build up the theoretical edifice. The theory must not be used to elucidate the concepts; never turn your eyes back. (Carrier (1990), p. 391)

In this section, we argue that both putative constitutive principles (1) and (2) can be decisively relaxed; in this way, a richer understanding of constructive axiomatics, constructivism, and their limits, is uncovered.

### 4.1 Theoretical versus intuitive constructive axiomatic approaches

Carrier's first constitutive principle of constructive axiomatisations—namely that axioms should be 'immediately amenable to experience'—leaves room for interpretation; what qualifies as 'immediately amenable to experience' is a function of context.

Does the EPS scheme display the first constitutive principle of constructive axiomatics? Well, how exactly are observations made according to EPS and how are they formalised into axioms? First of all, the most basic observation adhered to in the EPS scheme is that the 'world' consists of events; strictly speaking, EPS should have made this explicit as their first observational axiom.[41] Then, EPS note that there are different kinds of sets in event space $M$ corresponding to different histories of objects: one kind of set, which corresponds to the history of particles (simply called particles from then on), and another one, which corresponds to that of light rays (light rays from then on); again, the observation that there *are* particles and that there *are* light rays is suppressed as an explicit axiom of EPS. Once these two different types of event families are in place, one goes on to observationally characterise these two families of events (axioms $D_1$ and $D_2$) and ultimately—observing that these event families allow for coordinatising other events (axiom $D_3$)—one can start expressing observations about individual events (which will ultimately enable the observation of projective and conformal structure). At the general level, then, it seems fair to say that EPS bootstrap their own observational language, and with it, their theory of interest (GR), from the rather immediately accessible notions of events, particles and light rays, and one's immediate control over them.

It is important to stress, however, that EPS' axioms are usually not straightforward ascriptions of formal properties based on observations, but involve also non-empirical idealisations. Take EPS' initial axioms for establishing the differential structure: that a sequence of events associated to a particle through time ('worldline') is ascribed the status of a smooth manifold does not follow from 'immediate experience'; clearly, smoothness is a technical idealisation. Arguably, a naïve impression of continuity of the particle worldline (in the sense of a pencil line drawn in one go) may be something we can actually observe about particle worldlines. (Something analogous can be claimed about the smoothness/continuity of echos and messages.) Among the axioms on differential structure $D_1$-$D_4$, even the status of the more intricate axiom $D_3$ (postulating the existence of smooth radar charts on the set of events) seems to be in line with Carrier's characterisation of the first principle of constructivism: when freed from the obviously technical (albeit not harmless) assumption of smoothness, it seems that we do observe light rays to pave event space in the way stated by $L_1$—which, recall from Linnemann and Read (2021a), logically precedes $D_3$—and thus ultimately by $D_3$.[42] The axioms, however, also partly contain conventions that could not in principle be tested. For example, the assumption in axiom $D_4$ that light rays are smooth is not a mere idealisation, since *any* assumption about the one-way propagation of light is not amenable to empirical test (Salmon, 1977). Thus, one could say that, at least in some cases, the EPS axioms involve not merely idealisations, but also conventional assumptions.

---

[41]Arguably, EPS generally suppress any form of axioms that would amount to mere existence statements; rather, their axioms express observational attributions to already existing objects.

To sharpen the discussion of Carrier's first constitutive principle, let us introduce the term 'observation theory' for any theory whose empirical content is known and itself taken to be relatively well accessible. (We have already seen something of the notion of an observation theory in our above discussion of Einstein-Feigl completeness.) With this in mind, we then propose a crude distinction between (i) 'intuitive constructive axiomatics', a constructive approach which makes no recognisable recourse to any (physical) observation theory—all initial empirical statements are intuitively given and formalised, and more advanced statements expressed in a linearly self-erected observational theory[43]—and (ii) 'theoretical constructive axiomatics', a constructive approach which (at most) makes recourse to (physical) observation theories that are related to the phenomena independently from the target theory. The original EPS scheme is an instance of (i).

Now, as no data is genuinely theory-free (Stanford, 2021), the distinction between intuitive and theoretical constructive axiomatics is of course vague; strictly speaking, then, there can at best be a hierarchy of theoretical constructive axiomatics which differ from one another with respect to the degree of sophistication of their observational theories. This being said, it strikes us as fair and helpful to retain the bipartite distinction between intuitive and theoretical constructivist approaches (while recognising its limitations), and we will continue to make use of this terminology in the remainder of this paper. We give detailed examples of these two strands of constructive axiomatics in §2.3; before doing so, however, we consider the extent to which Carrier's second

---

[42]That said, note that we only observe this for a relatively sparse number of light rays in one small section of spacetime, so arguably there is something non-empirical going on in the generalisation of this axiom to all of spacetime.

[43]This is not to say that no non-physical theory, such as a logic or fields of mathematics, is available and used.

constitutive principle can be relaxed.

## 4.2 Linear versus iterative constructive axiomatic approaches

The other central pillar of constructive axiomatics for Carrier—*viz.*, the semantic postulate of linearity in the definition of theoretical terms, is also familiar from a more general project, namely Carnap's *Aufbau*. (See §5.9 for further discussion of the connections between the *Aufbau* and constructivism.) As (Carnap, 1967, §2) writes:

> To reduce [notion] $a$ to [notions] $b, c$ or to construct $a$ out of $b, c$ means to produce a general rule that indicates for each individual case how a statement about $a$ must be transformed in order to yield a statement about $b, c$. This rule of translation we call a construction rule or constructional definition (it has the form of a definition; cf. §38). By a constructional system we mean a step-by-step ordering of objects in such a way that the objects of each level are constructed from those of the lower levels. Because of the transitivity of reducibility, all objects of the constructional system are thus indirectly constructed from objects of the first level. These *basic objects* form the *basis* of the system.

In any case, a requirement of semantic linearity is at second glance unnecessarily restrictive: a proponent of constructive axiomatics might accept iterative definitions of terms instead of linear ones, at least under certain circumstances. Motivation in this direction can be drawn from the following paragraph from Carrier (who himself makes reference to similar ideas from Grünbaum):[44]

> One starts with an arbitrary geometry in the correction laws and determines physical geometry on its basis. The geometry obtained, however, does not generally coincide with the one used for the corrections. So in a second step, one employs the improved version of geometry to carry out the corrections and then repeats the whole procedure until an agreement is reached between the geometry entering the corrections and the geometry obtained by performing measurements with the accordingly corrected rods (compare Grünbaum 1973, p. 145). *If* this procedure of reciprocal adaptations indeed yields convergent results, the geometry that finally emerges is independent of the one used at the start and it is, furthermore, identical to the one obtained by exploring the coincidence behavior of transported rods in perturbation-free regions of space-time. So we can reasonably consider the resulting geometry as *the* geometry of space-time. (Carrier (1990), p. 385)

---

[44]It is worth noting that, for instance, Carrier maintains that Grünbaum's specific iterative approach faces difficulties; however, these potential problems will not matter for the general conceptual points which we seek to make in this subsection.

|  | **intuitive** | **theoretical** |
|---|---|---|
| **linear** | protophysics, original EPS | EPS with quantum matter |
| **iterative** | 'inventing temperature' | spin-2 bootstrap of GR |

Table 1: Varieties of constructivism

So in analogy, a constructivist—i.e., someone deserving of the title of a constructivist—could start with some previously known (empirically constructed) geometrical term that they are willing to leave 'free to flap in the breeze' in order to arrive at an even more convincing geometrical term; all that's assumed at the outset is immediately-given data (or data in terms of a *prima facie* acceptable observation theory), and that the initial geometrical proposal is in some weak sense based upon this data. For instance, in the flat-space spin-2 approach to GR, one starts out with a (local) Minkowski geometry together with a spin-2 field. From iterative correction procedures to the *dynamical* description of the spin-2 field, one then learns that the effectively resulting field replaces Minkowski spacetime as reference background geometry (or so the thought goes). Generally, a theoretical constructivism could allow for an iterative development that starts with and develops its observational theory in a self-correcting iterative rather than a straightforwardly linear fashion.

At this point, it is also worth connecting this notion of iterative (theoretical) constructivism with that of a 'completionism' *à la* Einstein and Feigl, as introduced by Carrier (1990), and already discussed above. At a very coarse-grained level: Einstein-Feigl completionism (EFC) requires a theory to contain its own observation theory. Although there are clearly connections here with iterative constructivism, the latter differs from EFC in the epistemological maxim: like the linear constructivist, the non-linear constructivist still starts out seeking to build up a theory from immediately accessible data through an accepted observational theory; but unlike the linear constructivist, she accepts that this might require a reciprocal approach. By contrast, the Einstein-Feigl completionist is motivated theoretically from the idea of minimal recourse to postulates external to the very theory under consideration in order to account for that theory's empirical predictions; for this they are, however, happy to assume the full theory from the start.

### 4.3 Examples

Given the above-introduced distinctions between intuitive and theoretical constructivism on the one hand, and linear and iterative constructivism on the other, we see now that there are four distinct strands of constructive axiomatics (see table 1). We rehearse here a paradigmatic example for each type:

**Linear-intuitive:** An example of a linear-intuitive constructivist project is that of protophysics (as conceived and fostered by Lorenzen and Janich and their respective students in the late 1960s to early 1980s). The goal is to give a systematic, non-circular (thus linear) account of basic physical quantities such as distance, time, and mass in terms of simple, craftman-like operations in one's everyday "lifeworld" ("Lebenswelt").[45] If successful, protophysics would provide a linearly derived basic theoretical relative to which empirical regularities can be formulated and judged on.[46] A second example of the linear-intuitive approach (and, indeed, axiomatisation) is of course that of EPS, introduced already in section 1.2, in which one begins with an intuitive observation theory in terms of light rays and freely falling particles, and constructs more sophisticated spacetime geometries (viz., Weyl geometries, and ultimately Lorentzian geometries) therefrom.

**Linear-theoretical:** For an example of a linear-theoretical constructivist approach, consider the quantum mechanical generalisations of EPS developed by Lämmerzahl (1998). In these approaches, one begins with quantum mechanical waves (*qua* solutions of the Schrödinger equation), and again constructs therefrom spacetime geometric notions, and so forth. Although conceptually this approach is transparently linear, and in the same spirit as the original work of EPS, a key difference is that the initial observation theory in this latter case—viz., a theory of quantum mechanical waves—is substantially less 'intuitive' than that of the original EPS approach.[47] For this reason, it is reasonable to count it a case of linear-theoretical constructive axiomatisation.

**Iterative-intuitive:** An example of an iterative-intuitive constructivist project is that of the development of the field of thermometry, as presented in the book-length study by Chang (2004). According to this, one begins with elementary empirical observations, but then undertakes a "self-improving spiral of quantification—starting with sensations, going through ordinal thermoscopes, and finally arriving at numerical thermometers" (Chang, 2004, p. 221). Such cases are also discussed—admittedly somewhat more abstractly—in (Read and Møller-Nielsen, 2020), in which it is argued that the relation between theory and observations can be understood within the framework of 'hermeneutic circle style reasoning' (cf. Van Fraassen et al. (1980)), in which one begins with a basic observation theory, and on the basis of said theory constructs some more sophisticated theoretical edifice; that very edifice may, however, lead one to revise one's observation theory, at which point the process repeats.

---

[45]In fact, prior to grounding basic physical quantities, geometric notions and arguably even logical concepts have to be grounded linearly in everyday practice.

[46]The collection (Böhme, 1976) and the *Philosophia Naturalis* special issue on protophysics, edited by Janich and Tetens (1985), are good starting points to the literature. Major works include the proto-geometric accounts by Inhetveen (1983), Lorenzen (2016), and Janich (1997), and the proto-physical account of time by Janich (2012).

[47]To some extent, it's hard to imagine an observation theory *less* intuitive, in light of the quantum mechanical measurement problem! In fact, it is even unclear that the Lämmerzahl construction really satisfies the desideratum of constructiveness above—that the (physical) observation theory (should be) strictly closer to the phenomena than the target theory.

**Iterative-theoretical:** An example of an iterative-theoretical constructivist project from physics practice is the famous spin-2 bootstrap of the Einstein field equations: the second-rank Lorentz-invariant tensor field $h_{\mu\nu}$ is postulated in a flat spacetime $\eta_{\mu\nu}$ such that it obeys the second order equation of motion usually associated to linearised gravity (called the 'Fierz-Pauli equation'). The central idea then is to allow for a sourcing of the Fierz-Pauli equation from matter energy-momentum content. As the resulting equation is, however, found to be inconsistent (hitting both sides of the matter-sourced Fierz-Pauli equation of motion with a divergence operator leaves the Fierz-Pauli part equal to zero while the divergence of the matter energy-momentum tensor is unequal to zero; if it were equal to zero as well, there would be no interaction with the Fierz-Pauli field $h$ to begin with), one starts, as a remedy, to take into account the energy-momentum tensor associated to the $h$-field itself. Contrivedly, adding an energy-momentum tensor for $h$ from the given equation of motion generates a new equation of motion for $h$ whose energy-momentum tensor contribution to $h$ has to be taken into account as well so that an iterative relation is generated. In the limit of infinite iterations, the Einstein field equations for a composite field $g_{\mu\nu} := \eta_{\mu\nu} + h_{\mu\nu}$ where $\eta$ is the Minkowski metric, are claimed to arise.[48]

## 5 Constructive axiomatics in context

Having now (a) introduced some of the history of constructive axiomatics (§1), (b) explained the connections between the EPS axiomatisation and Weyl's theorem, and presented some attempts to fill holes in the original EPS approach (§2), (c) considered the relations between constructive axiomatics and chronometry (§3), and (d) classified different varieties of constructive axiomatics (§4), we turn now to considering the connections between constructive axiomatics and various other projects and research programmes in the foundations of spacetime theories.

### 5.1 Other constructive axiomatisations of spacetime theories

The literature includes various other constructive axiomatisations of spacetime theory (see Castagnino (1968); Hayashi and Shirafuji (1977); Hehl and Obukhov (2006); Majer and Schmidt (1994); Schelb (1996b); Schröter (1988); Schröter and Schelb (1992b,a)) that are, albeit less known, very similar in spirit to EPS. On the other side of the spectrum, there are also straightforwardly deductive axiomatisations (see Andréka et al. (2002); Benda (2008); Bunge (1967); Cocco and Babic (2021); Covarrubias (1993); Mould (1959) for some examples). Interestingly though, certain accounts cannot be easily filed as constructive or deductive. An example in this direction is for instance the account by Hardy.

---

[48]Despite the seeming appearance even of a derivation, the inference suffers from several ambiguities. See, for instance, Padmanabhan (2008) and Baker et al. (2022). A central criticism concerns ambiguities in setting up the consistency condition due to ambiguities in how to calculate the energy-momentum tensor for $h$.

Hardy (2018) argues for a broadly constructive approach to spacetime, not merely as a way of explaining classical spacetime but also as a promising route towards quantum gravity (see Adlam et al. (2022) for further discussion). He suggests that Einstein's original route to GR can be understood as a 'construction' based on seven principles (which single out the general framework for spacetime theories, i.e. the kinematics) and three additional elements (which single out general relativity, i.e. the dynamics). We may therefore compare the approach of Hardy (via Einstein) for constructing the kinematics to the EPS construction. In the schema of §4, Hardy's approach is clearly theoretical rather than intuitive since the seven principles don't describe immediate empirical facts but rather prescribe steps in setting up a coordinate system and putting fields on it—for example, one of the principles is 'there is no global inertial frame,' which is not something that any local observer could hope to establish empirically (we might find local phenomena which we regard as evidence for the nonexistence of a global inertial frame, but we cannot make global observations and thus we cannot directly observe the existence or nonexistence of a global frame; and moreover it is unclear that 'nonexistence' can be subject to direct observation even in more ordinary cases). It is also structurally linear, in the sense that it involves no iterative steps, although one might contend that some of the principles seem to rely on specific knowledge about the target theory in a way which makes the approach somewhat less linear—for example, one of the principles asserts that the theory should be defined in terms of local tensor fields based on the tangent space, which seems difficult to justify in terms which don't take account of background knowledge about general relativity.

Hardy goes on to propose a tentative 'constructive' approach to quantum gravity, with some of the details yet to be filled in. As in the case of Hardy's construction for GR, this approach is clearly theoretical rather than intuitive. Indeed, the axioms are necessarily very far removed from any possible empirical observations. For example, one axiom postulates the existence of 'indefinite causal structure,' which does not correspond to any known observation—-and in fact it is quite difficult even to come up with a hypothetical empirical observation which could be regarded as instantiating indefinite causal structure. Given this feature, it's unclear whether we would want to regard Hardy's approach to quantum gravity as 'constructive' in the sense of EPS—perhaps it should simply be taken as a principle theory with non-empirical principles.[49]

---

[49] For more on constructive approaches to quantum gravity, see Adlam et al. (2022).

## 5.2 Relationship to constructive mathematics

The constructivist undertaking of EPS (and related programs mentioned above) shares motivations with constructive mathematics. Both approaches regard non-constructivist procedures as being in some sense epistemically opaque. Some familiarity with constructive mathematics is therefore instructive in order to appreciate fully the epistemological background assumptions at play in any constructivist effort.

Constructive mathematics is agnostic about the law of excluded middle—i.e., '$\forall p \in$ the set of Propositions: $p \vee \neg p$'—and thus dismisses it for all practical purposes, including as a proof technique. This scepticism towards the law of excluded middle is tantamount to scepticism towards proofs by contradiction of the form $\neg\neg p \to p$ (basically, a reformulation of the law of extended middle). For the latter, it is quite clear why its assumption could be problematic: the proof-by-contradiction is indirect and thus opaque; rather, a proof should (a constructivist about mathematics would maintain) explicitly 'construct' what is supposed to be shown. Note that an actual dismissal of the law of excluded middle would also mean the dismissal of the axiom of choice (as the latter implies the former: see (Bauer, 2017) for a simple demonstration).

Now, if the usual repertoire of proof techniques becomes impoverished to the above degree, then it is not clear how much of standard mathematics (or something close to it) can be reproduced.[50] It was Bishop (1967) who, however, for the first time demonstrated that a mathematical field—in his case that of analysis—could be given a viable constructivist reformulation. It is worth stressing that constructivism *à la* Bishop is of an epistemic character—it derives from a skepticism towards methodology. Brouwer's constructivism, known as 'intuitionism'—the first form of mathematical constructivism—, arguably shares these concerns but the ultimate motivation is ontological: mathematics, a mind-dependent affair, needs to be studied by seeing how it is built up on a minimal ontology of intuitive statements (axioms). (See Bridges et al. (2022).)

If we now look at the case of spacetime constructivism, we see that there are clear parallels to epistemic mathematical constructivism *à la* Bishop: rather than accepting the existence of some mathematical structure as empirically relevant just because it allows for deducing empirically testable results, basic experiences should—admittedly under some idealisations—allow for constructing the theoretical structure of interest. (The aforementioned research program of protophysics has been presented by its proponents (see e.g. Janich (2012)) as an extension of constructive mathematics into the regime of physics. See our above discussion of protophysics as an intuitive constructivist approach to physics.) Also worthy of mention here is that a significant advantage of constructive mathematics is that the proofs are not merely existence or uniqueness proofs, but rather provide instructions for *how to construct the object in question*, which is sometimes a very useful and informative thing to be able to do. So ar-guably the motivation is not just epistemic but also pragmatic: here, too, there is an analogy with constructivism in physics, in the sense that the latter reveals the basic 'constituents' of the objects under consideration (e.g., projective and conformal structure composing a Lorentzian metric field), and how said constituents might also be arranged in different ways, in order to arrive at different structures (e.g., Finsler geometries, as we have seen).

---

[50]Hilbert's lines of complaints in this direction are famous: "Taking the principle of excluded middle from the mathematician would be the same, say, as proscribing the telescope to the astronomer or to the boxer the use of his fists." (Hilbert, 1927)

## 5.3  Constructivism and theory formulations

Famously, Einstein in 1919 drew a distinction between 'principle theories' and 'constructive theories'. Here is what he wrote:

> We can distinguish various kinds of theories in physics. Most of them are constructive. They attempt to build up a picture of the more complex phenomena out of the materials of a relatively simple formal scheme from which they start out. Thus the kinetic theory of gases seeks to reduce mechanical, thermal, and diffusional processes to movements of molecules—i.e., to build them up out of the hypothesis of molecular motion. When we say that we have succeeded in understanding a group of natural processes, we invariably mean that a constructive theory has been found which covers the processes in question. Along with this most important class of theories there exists a second, which I will call "principle theories." These employ the analytic, not the synthetic, method. The elements which form their basis and starting-point are not hypothetically constructed but empirically discovered ones, general characteristics of natural processes, principles that give rise to mathematically formulated criteria which the separate processes or the theoretical representations of them have to satisfy. Thus the science of thermodynamics seeks by analytical means to deduce necessary conditions, which separate events have to satisfy, from the universally experienced fact that perpetual motion is impossible. (Einstein, 1919)

Alongside phenomenological thermodynamics, Einstein identified his 1905 theory of special relativity as a principle theory. Now, while much has been written on the explanatory merits of constructive theories over principle theories (see e.g. Read (2020b)), what we wish to point out here is that there is a sense in which constructive axiomatisations—at least, intuitive constructive axiomatisations—are more akin to *principle* theories, rather than constructive theories. The reason is that they precisely seek to build up the theories in question from empirically well-grounded axioms.[51]

---

[51]From a systematic point of view, the principle-constructivist theory distinction is first of all an epistemic distinction in the sense of how we can learn about a theory of interest. Compare this to the framework vs. concrete dynamics distinction (as for instance pushed by Benitez (2019)), which is trying to distinguish the levels in abstraction between theoretical expressions and theories as

Does this cast any aspersions over the constructivist project? In our view, it does not, for attempting to set any given theory on firm empirical footing is perfectly consistent with attempting to identifying some deeper, physical explanation as to why that theory holds. Harvey Brown (p.c.) is suspicious of constructivist approaches such as EPS as a result of their being "too operational" (see below for more on how to make sense of this claim)—however, in our view, such suspicion is unwarranted, once one recognises the compatibility of the two approaches as per the above. It also deserves to be recognised that axiomatic approaches such as EPS seem to afford a means by which Brown's preferred programme of the ontological reduction of spatiotemporal structure to material bodies can proceed—in this sense, he should also recognise the advantages of the approach (see below for further discussion).[52]

Moreover, the constructivist project may also be seen in light of a broader movement in physics towards taking principle theories more seriously. For example, Grinbaum (2007) argues that operational axiomatisations of quantum theory should be seen as principle theories and argues that increasing interest among physicsts in operational theories shows that principle theories are now being taken seriously as an end in and of themselves, rather than merely a step on the road to a constructive theory. Grinbaum (2017) interprets this trend as a move towards a form of idealism in which physics is understood to be primarily about language; Adlam (2022), on the other hand, interprets this trend within the realist tradition as a move away from object-oriented realism in favour of a more structural realism. Both approaches agree that these operational axiomatisations have some advantages over the traditional constructive approach insofar as they offer a way of making sense of a theory which requires very few commitments to unobservable or theoretical entities— something which seems particularly valuable in the case of a theory like quantum mechanics where the ontology of the theory is not at all transparent. More generally, similar claims may be made for other constructivist projects, including EPS: these 'principle theories' have value in their own right, and need not always be regarded as inferior to the complementary constructive theories.

The particular relevance of the EPS axiomatisation in this context is that it emphasizes the fact that general relativity can be regarded as a principle theory, in much the same sense as SR is often acknowledged to be a principle theory.[53] Note that both SR and GR, *qua* principle theories, serve as *constraints* on possible dynamics for material fields (in the case of SR, that all material fields must have dynamics governed by Poincaré invariant laws; in the case of GR, that all material fields couple to a dynamical metric field). This is to be contrasted with

---

such. In particular, a theory such as special relativity can be a principle theory, i.e., in the sense that it can be motivated phenomenologically, while at the same time star as the fundamental framework in a bottom-up model.

[52]We thank Chris Smeenk for discussion on these points.

[53]The viewpoint of GR as a principle theory seems also in line with its thermodynamic-hydrodynamic interpretations, as put forward by Jacobson (1995), Padmanabhan (2012, 2011), and also Hu (1999). Furthermore, even at the level of dynamics, it seems that there is phenomenological input into GR, say via the rather coarse-grained coupling in the Einstein equations of the metric field to matter in terms of energy-momentum alone.

thermodynamics, which is a purely phenomenological principle theory—there is no remaining freedom to postulate different dynamics consistent with the principles of the principle theory.

## 5.4 Constructivism and empirical interpretation

The EPS scheme (or constructive axiomatics more generally) can be seen as one specific way to address the interpretational problem for GR—i.e., the question of how to link its formalism to the empirical data and thus, in a minimal sense, to the world (see Carrier (2018); Dewar et al. (2022)).[54] Whereas GR is standardly interpreted through fixed (often chronometric—see above) correspondence principles (see Synge (1959) and Malament (2012) for explications in this direction), the constructive axiomatic approach *à la* EPS incrementally generates ever more fine-grained correspondences, starting only with a basic representational ontology in terms of a set of events of which particles and light rays are supposed to be subsets.

Compare, for instance, Synge's presentation of the clock hypothesis as a correspondence principle to the way in which correspondences are built up by EPS. The clock hypothesis of general relativity equates the worldline interval length of a point particle trajectory to the actual proper time experienced by that particle; it becomes an interpretational principle at latest once the particle is understood as a stand-in for an idealised clock. By contrast, constructive approach starts out by linking unspecified objects (set of events, particles and light rays) to the world which get rendered in a more and more detailed fashion via brute ascriptions (particles, for instance, are ascribed the status of smooth one-dimensional manifolds; furthermore, echos and messages defined between them are ascribed to be smooth, etc.). In the constructive approach, one thus dresses up an unspecified theoretical object which just acts, in a sense, as an 'anchoring' of the object in the world (say a set of events corresponding to a real-world 'particle'), with ever more detailed formal description. This is in contrast to the standard chronometric coordinative definitions which associate 'finished products' in the theory to the world.

The lesson we learn is that, even though constructive axiomatics does not evade the need for correspondence principles *per se*, it can take the weight from correspondence principles by reducing them to a much weaker subset. After all, the usual concern with correspondence principles is that they are disturbingly brute stipulations—an impression which is, arguably, successfully weakened in constructive axiomatics.

Another (quite distinct) approach to the empirical interpretation of GR begins with the *local* validity of special relativity, and bootstraps from this via certain additional principles to the kinematical structure of general relativity (these ideas are discussed by Brown and Read (2016), and more explicitly by Hetzroni and Read (2022)). This is in the spirit of a point raised by Lehmkuhl (2021), that the 'strong equivalence principle' (by which here is meant the assumption of the local validity of SR) can lead to the "trickling up" of the interpretation of SR to GR. (Such ideas are also suggested in (Brown, 2005, ch. 9).) Immediately, one can see that such approaches are more theoretical (in our sense used above) than e.g. EPS, for they *begin* with the full theoretical edifice of special relativity. On the other hand, arguably one advantage of said approaches is that they thereby subsume *all* the empirical evidence which led to special relativity, before attempting to bootstrap to some new theory (namely, the general theory), rather than beginning with a very impoverished (and, for that reason, some might argue less physical) set of starting axioms. Our conjecture is that it is this latter aspect which leads Brown to prefer such approaches over the EPS construction (as mentioned above).

---

[54]The question of whether general relativity needs an interpretation is also taken up by Belot (1996), Curiel (2009), and Linnemann (2021).

## 5.5 Constructivism and operationalism

The EPS scheme has a strong operationalist flavour. It is important to distinguish, then, as to whether operationalism is meant here as a theory of *meaning*, or rather in a moderate fashion which stresses the need for operationalist *analysis*, rather than the absolute necessity of operationalist *definitions*. We take it that it is indeed not necessarily the case that the constructivist considers a theory-first approach as meaningless unless linked to the world by some operational interpretative rule. More likely, they will simply consider it less satisfactory than the epistemically less opaque route of linking the theoretical structures directly to experience. Given that operationalism as a theory of meaning has become outdated (Chang, 2009), we will consider the moderate stance in the following.

To what extent does the EPS scheme then succeed with operationalist analyses? As we have seen above, the EPS scheme is limited to local neighbourhoods at least for non-static spacetimes. Even in the case of static spacetimes, any form of sensible tracking requires encodings within the signal (to showcase emission and receival times); from an operationalist point of view, the EPS scheme is in this respect really just schematic. In any case, it seems fair to see in the EPS scheme succeeds as a form of a conceptual operationalism: one can lead back spacetime to manipulable operations with particle and light trajectories at least as a form of helpful mental picture.[55]

But more than that, the EPS approach does seem to provide a scheme which, once fleshed out through appropriate choice of particle and signalling items, promises to be physically realisable.[56] Notably, quantum EPS—see the intro-duction of (Adlam et al., 2022)—may indeed not be regarded as an experimentally realisable scheme; in this sense, it is less operational (in a practical sense) than the classical EPS scheme.

---

[55]Cf. Bridgeman's idea of mental, verbal and/or paper-and-pencil operationalism—as opposed to that of a 'laboratory operationalism'. As Chang (2009) notes, Bridgman lamented that it was the "most widespread misconception with regard to the operational technique" to think that it demanded that all concepts in physics must find their meaning only in terms of physical operations in the laboratory (Bridgman, 1938).

[56]The work of Audretsch and Lämmerzahl (1991) does replace the idea of particle by that of matter wave, but note that thereby just one yet rather theoretical concept is swapped for another one; even on this account, then, matter wave and light rays need to fleshed out further by the experimenter. In a sense, none of this is of course news but rather part of the business of experimental

## 5.6 Connections with geodesic theorems

In a typical textbook presentation of general relativity, the geodesic motion of test bodies is assumed—see e.g. (Malament, 2012, ch. 2). However, since (at least—see below) the work of Geroch and Jang (1975), authors have sought to *derive* the geodesic motion of small bodies from the Einstein equations in general relativity. (Actually, work to prove geodesic motion general relativity—albeit of a very different kind to that of Geroch and Jang (1975)—goes back to Einstein and Grommer: see Tamir (2012) and Lehmkuhl (2017) for discussion.) Successor papers to (Geroch and Jang, 1975) include (Ehlers and Geroch, 2004)—in which back-reaction between the small body and the metric field is accounted for—and (Geroch and Weatherall, 2018)—which uses the machinery of 'distributions' in order to derive in addition that small *massless* bodies follow null geodesics. Insofar as one might then use these motions of small bodies in order to reconstruct the metric field via Weyl's theorem (Weyl, 1921), one might think that such geodesic theorems secure full access to the metrical structure of spacetime (something along these lines is suggested by Read (2020a)).

This reasoning, however, is somewhat confused. All geodesic theorems of the kind discussed above are proved from within the context of the *completed* theory of general relativity—thus, they have very little to do with the programme of *constructive* axiomatics *à la* EPS. That being said, having (from within the completed theory) derived such motions, one could of course in turn apply the EPS machinery to these paths in order to re-derive the original metrical structure from which one began, as a kind of consistency check. Aside from this, however, it's not entirely natural—*pace* Read (2020a)—to situate such geodesic theorems alongside EPS, when the former begin with the very theory (the kinematics of) which EPS is designed to recover.

## 5.7 Relationship to conventionalism

Within the epistemology of geometry (on which see e.g. (Dewar et al., 2022) for an introduction), constructive axiomatics (on which, as should by now be extremely clear, the kinematics structure of a theory is to be built up from elementary axioms which are supposed to have direct empirical significance) is sometimes set apart from conventionalism (according to which the structure of space and time is a conventional matter, which must be chosen on the basis of extra-empirical considerations), the latter being most famously associated with Poincaré (1902). What our discussions here make clear, though, is that constructive axiomatics *à la* EPS in fact goes hand-in-hand with conventionalism. For example: when EPS rule out Finsler geometries or torsionful ge-

---

physics when carrying out theoretical proposals in the concrete.

ometries, they essentially do so by conventional stipulation. Moreover, we've already seen that some conventional stipulations (e.g. regarding freely-falling particles) are retained in EPS, even on Coleman and Korté's modified version of the scheme.[57]

One central issue regarding conventionalism pertains to the extent to which the theory under consideration depends upon the assumptions made about the observation theory in use. As already insinuated, at a general level, independent of the specific case of the EPS scheme, there is a sense in which the linear constructivist approach can never fully succeed in building up theoretical structure 'from scratch': there are always conventional choices to be made, and the constructivist thereby only 'feigns' not to know what she is genuinely after. For instance, as Carrier (1990) notes, "the EPS-scheme does not relieve us from the need to decide about the presence or absence of universal forces".

## 5.8 Constructivism and the dynamical approach

The dynamical approach to spacetime theories, promulgated by Brown (2005); Brown and Pooley (2001, 2006), and summarised recently by Brown and Read (2021), is an alternative to constructive axiomatics as an account of why the structure of spacetime is what it is (rather than otherwise)—this contrast was drawn recently by Dewar et al. (2022). One of the core tenants of the dynamical approach, at least in the context of theories with fixed spacetime structure such as special relativity or Newtonian gravity, is that spacetime structure *just is* a codification of the symmetries of the dynamical equations governing material fields, so that ultimately the nature of spacetime is to be explained by appeal to features of the dynamical equations of motion.

Now, the constructive axiomatic approach may initially seem very different in spirit to the dynamical approach, insofar as the EPS construction (say) proceeds entirely from empirically observed motions without invoking any equations of motion at all (cf. our discussion in the final paragraph of §5.4). But following Anandan (1997), it is possible to see connections between the approaches—because after all, symmetry groups must also have something to do with the empirically observed motions, in which case it must be possible to extract the symmetry groups from those motions.[58]

In particular, Anandan notes that in a sufficiently small region, the affine structure as defined early on in the EPS construction has as its symmetry group the affine group generated by the general linear transformations and translations in a 4-dimensional real vector space. This affine group has as subgroups the inhomogeneous Galilei group and the Poincaré group—the former corre-sponding to non-relativistic physics and the latter to relativistic physics. Moreover, focusing on the Poincaré group, it in turn has as subgroups the translational subgroup and the Lorentz subgroup; the former acting on a small region around a given point determines the projective structure, while the latter leaves invariant the null cone at each point and hence dictates the conformal structure, and the relationship between these two spacetime structures can be understood in terms of the relationships between these two subgroups of the Poincaré group. Thus the route taken by EPS to arrive at projective and conformal structure from affine structure can be abstracted in terms of the symmetry group of the affine structure, such that the derivation can now be understood in something like the language of the dynamical approach. Thus, there is in fact a reasonably close correspondence between the dynamical understanding of spacetime in terms of symmetries and the constructive approach in terms of the behaviour of particles and rays. Indeed, in a sense the explanations of relativity offered by the dynamical approach can be regarded as simply a reformulation of the explanations offered by the constructive approach, arising naturally when we abstract the symmetry groups away from the behaviour of test particles and arrive at spacetime on that basis. (This being said, a proponent of the dynamical approach might still complain that the EPS methodology *qua* route into GR does not begin sufficiently rich theoretical structure, and in this sense remains 'too operational'—recall again our discussion in §5.4.)

---

[57]In this respect, constructive axiomatics is distinct from protophysics, the proponents of which claim to be able to circumvent all matters of conventionalism: see (Dewar et al., 2022).

[58]We recognise that the symmetries of solutions of equations need not be the symmetries of those equations themselves—see (Read and Cheng, 2022) for a discussion of this point in something like this context. Nevertheless, it is surely true that the behaviour of material bodies, described by a particular solution to a particular equation, must have *something* to do with the symmetries of that equation, for otherwise the equation would not describe accurately the target phenomena.

Moreover, recall that in the original EPS construction, in order to arrive at Riemannian geometry from Weyl geometry, EPS must invoke a somewhat *ad hoc* requirement: the stipulation that there should be no 'second-clock effect'; i.e., that parallel-transported vectors should not change in length. But Anandan contends that a less *ad hoc* approach emerges naturally within the dynamical picture if particles are replaced with quantum matter ones, subject to the requirement that the waves approach particle-like behaviour in the geometric optical limit.[59] For quantum waves have a natural frequency given by $mc^2 = \hbar\omega$, and thus the phase operator of such a wave can be used as a clock. In particular, we may imagine two such waves travelling from a common origin to a common destination; the metric along each path is determined by the Casimir operator $m^2$, but meanwhile the gravitational phase operator which generates the evolution along the path commutes with the Casimir operator, and therefore the Casimir operator remains the same as it is transported along each path, which means that the clocks must agree again when they meet. Thus, in the quantum context the 'no second-clock effect' axiom can be regarded as following directly from fundamental dynamical symmetries.

This is interesting for several reasons. First, it gives us a tantalising glimpse of a deep underlying connection between quantum mechanics and classical general relativity: perhaps the reason EPS could not do without the *ad hoc* 'second-clock' postulate is because in order to fully understand and rationalise the nature of classical spacetime one must take into account its quantum un-derpinnings.[60] Second, it suggests that, *contra* the approach of EPS, it may in fact be necessary to invoke some dynamical considerations in order to arrive at the structure of spacetime: it seems that kinematics and dynamics are more closely intertwined than one may initially have suspected, and perhaps a complete understanding of the nature of spacetime depends on an understanding of the relationship between them. Of course, this latter lesson is in the spirit of Brown (2005).

---

[59]Cf. Audretsch and Lämmerzahl (1991).

### 5.9 Constructivism and *Der Aufbau*

As already alluded to by borrowing Carnap's definition of 'construction' for our own purposes, EPS' undertaking resembles Carnap's project in *Der logische Aufbau der Welt*. In general, it seems worthwhile to relate wider-ranging constructivist projects *à la* Carnap to domain-specific (i.e., physics-specific) constructivist accounts such as EPS: this task we undertake in the present subsection.

In *Der Aufbau*, Carnap considers hierarchical layered systems of statements—'constitution systems' (traditionally translated as 'constructions'!)—in which the vocabulary for statements in higher layers can be defined in terms of the vocabulary for statements in the next-lower layer and in which the lowest layer's vocabulary is called a 'basis'. The ambitious claim is then that the world as such can be expressed in terms of various constitution systems, say by using a physical basis, i.e. basic physical facts, a hetero-psychological basis, i.e., basic facts of observers' experience, or an auto-psychological basis, i.e., basic facts about one observer's very own experience. Importantly, Carnap stresses that the purpose of such reconstructions of the world in terms of constitution systems need not be epistemic; and that even though he himself takes the auto-psychological basis to be most fruitful when there is an epistemic interest behind setting up a constitution system, he explicitly leaves open whether such choice in the epistemic context can be overturned in the future.[61]

Despite such explicit qualifications, Carnap's work has in particular in the English-speaking world been read—or, rather, in lack of a proper translation, been reported by Ayer, Quine and others—as a project of reductionism in the British empiricists' tradition. However, a second major reading of Carnap's motivation has over time emerged—in the English-speaking literature in particular spearheaded by Friedman (1999)—which puts Carnap's intention much more into context to his phenomenological and neo-Kantian influences in Germany at the time of writing.[62] For Friedman, for instance, the starting motivation to *Der Aufbau*, with its *de facto* one-sided commitment to the auto-psychological basis, indeed lies in an attempt to bridge the gap between own's basic phenomenological impressions on the one hand and the objective physical world on the other, for the sake of structuring and solidifying the status of the phenomenological; it concerns the status of the phenomenological not of the outside world, or—more in the neo-Kantian spirit—the gap between the subject and the world as such. But as Friedman proceeds to argue, this original motivation ultimately gave way to the thought that the constitution systems can be read in metaphysical or epistemic dimensions at will as long as its content in a narrow sense is seen (and only seen) in structural relationships; the adherence to this predecessor of the tolerance principle then ultimately makes the phenomenological as well as the neo-Kantian motivation only one out of several possible readings (and also leaves room for the reading in terms of empiricist reductionism as an additional option).[63]

---

[60]This we take to be in the spirit of our Part III: see Adlam et al. (2022).

[61]Advances in neuropsychology (or so he speculates) might fancy a more naturalised approach again, and taking its basic vocabulary as a basis. See (Leitgeb and Carus, 2022, Supplement A).

[62]For another reading, see, for instance, Pincock (2005).

Now, *Der Aufbau*—no matter in which reading—is widely considered a failure. However, Leitgeb (2011)—who has set out to weaken and smoothen its claims to arrive at a more viable version—argues for reviving *Der Aufbau*'s project *qua* epistemological agenda in two different ways: (i) partly responding to the original empiricist motivations, an updated *Aufbau* can demonstrate a relevant part of (albeit not exhaust) the meaning of expressions. As Leitgeb details: "if experience is understood in terms of a subjective basis that is relativized to a particular cognitive agent, then the so-determined empirical meanings may be considered to be among the internalist meaning components of linguistic expressions—the meaning components that are 'in' this agent's mind—which are additional to externalist (referential) ones." (p. 270) So, accepting with the mainstream that what is empirically accessible does not exhaust the meaning of theoretical terms, constructive axiomatics allows for working out the extent to which the individual observer can remark about the totality of meaning of theoretical terms (for the internal-external meaning distinction, cf. Putnam (1981)). Secondly, partly responding to Carnap's original neo-Kantian motivation (at least on Friedman's reading), a new *Aufbau* may still be used to "fill the gap between subjective experience and the intersubjective basis of scientific theories. After the protocol sentence debate in the early 1930s, philosophers of science more or less decided to conceive of the observational basis of science as being intersubjective right from the start; observation terms and observation sentences were meant to refer to observable real-world objects and to their observable space-time properties." (p. 270) In the case of EPS, one might then find justification in the project in that it helps to connect various observer viewpoints (you with your particles and light rays; me with mine), solidifying their viewpoints as parts of a single scientific outlook (that of GR).

Moreover, one might argue that a revived EPS constructivist project can be motivated for similar reasons: thanks to an EPS-like scheme, one learns about the operational content of projective/conformal geometry; and thanks to EPS, one is able to link the observer explicitly into the otherwise highly theoretical structure of the full kinematical picture. Notably, this second point is much stronger than a mere schematisation of the observer within a given theoretical structure (as Curiel (2019), for instance, arguably likes to think of it)—one reduces adherence to correspondence rules to a significant degree (in analogy to how relying on one's own protocol sentences, rather than relying on just anyone's protocol sentences, reduces adherence to correspondence rules to a significant degree).

---

[63]See (Leitgeb and Carus, 2022, Supplement A) for a detailed account of the *Aufbau*'s reception.

# 6 Towards a restricted foundationalism?

We have noted before that a constructive-constructivist project *à la* EPS has some programmatic dependence (if not circularity) on the goal of its construction—*viz.*, GR; arguably, it is thus not totally constructivist. In addition, it would be naïve to assume that the full content of a theory can be grasped constructively, i.e., from a non-theoretical or substantially less theoretical basis—we know by now, not just through the likes of Duhem, Neurath, Hanson and Quine, that the theoretical is *generally* not exhausted by the empirical, and that (some sort of) coherentism has won over foundationalism.[64] In this section, we want to elaborate on what precise 'restricted' meanings can be given to constructive-constructivist takes on physical theories, despite the general consensus that empirical foundationalism—to which constructive-constructivist programs first of all seem to aspire—fails.

A central use of a 'merely' restricted empirical foundationalism lies in bridging conceptual gaps (rather than seeking ultimate justification in the empirical). We make this point concrete in (Adlam et al., 2022) by showing how an EPS-like scheme can be used to provide novel understanding of 'quantum spacetime', namely in terms of quantum signals: issues of quantum superposition of spacetimes (including the possibility of quantum diffeormophisms[65]) are led back to issues of how to think of quantum signals; this is advantageous to the extent that we have better intuitions for sensible options of how superposed signals can behave than for how superposed spacetimes can behave. Arguably, such a point for restricted foundationalism could already be made by reference to EPS—but it is better made by reference to quantum EPS: all achievements by EPS on the conceptual level (say how to understand spacetime from the observer's eye) might be discarded as mere reformulations of previous insights (although we ourselves do not think so); in the case of quantum EPS, however, it is evident that spacetime superpositions have issues and that we make a leap forward in understanding these issues (or even identifying some of them to begin with) thanks to the constructive-constructivist approach.

For completeness, it is also worth pointing out that a constructivist mindset can usefully be put to action even across theories rather than between a theory and (parts of) its observational basis—and thus also independently of

---

[64]See also the work of Carrier (1990), in which the shortcomings of EPS *vis-à-vis* the demand for a theory to include its own observation theory (Einstein-Feigl completeness again) are discussed.

[65]Relative to the wavefunction expressing the superposition of spacetimes, a 'quantum diffeomorphism' is the simultaneous application of individual diffeomorphisms to the different branches of the wavefunction (and the associated manifolds). The notion is discussed further in (Adlam et al., 2022).

whether the construction basis is constructive in the narrow sense or that of Reichenbach or rather just admissible in some wider fashion. For this, consider the issue of spacetime emergence in GR from theories of quantum gravity: let us agree for a moment that quantum gravity approaches are indeed non-spatiotemporal in some relevant sense (see Le Bihan and Linnemann (2019); Linnemann (2021) for a dissenting view, and see Jaksland and Salimkhani (2021) for a pertinent critique of the loose usage of the words 'spacetime' and 'emergence' in spacetime emergence claims). The question that has kept people busy in the philosophy of quantum gravity community, then, is: how can spatiotemporal structure arise from non-spatiotemporal structure? Could it not be that the whole empirical success of standard physics—based on measurements in space and time—is undermined as long as it is not clear that spacetime exists fundamentally? (Huggett and Wüthrich, 2013)

Spacetime functionalists (e.g. Lam and Wüthrich (2018, 2021); Huggett and Wuthrich (2021)) argue that the physical salience of these (presumably) non-spatiotemporal structures can be vindicated by showing how that structure manages to functionally realize spatiotemporal roles. But note: a functional reduction (as with any reduction) first of all explains the reduced by the reducing—and so the non-spatiotemporal from the spatiotemporal—and not *vice versa*.

Admittedly, the proponents of spacetime functionalism claim that the vindication of physical salience through a (functional) reductionist scheme runs both ways. As an intuition pump, Huggett and Wuthrich (2021), for instance, bring up the example of the physicalist's reduction of mental states: they take it to be immediately plausible that the physicalist's functional reduction of pain in terms of brain state configurations can not only be read to strengthen the status of pain in the face of physicalist's concerns. Rather, the physicalist's reduction can also strengthen the status of brain states in the face of phenomenalist's worries. This leaves room for concern though, as the intuition here is a very different one from that of the constructivist. From the constructivist point of view, a physicalist reduction of pain bridges the gap from a physicalist understanding towards a phenomenalist understanding (the physicalist can understand, i.e., can model the phenomenalist in her language). However, only a phenomenalist reduction of mental states would decisively bridge the gap *given* a phenomenalist basis to a physicalist world view (the phenomenalist would understand, i.e., could model—at least to some extent—the physicalist's renderings in her language).

The constructivist who criticizes spacetime functionalism at the same time can make the positive point that the conceptual gap issue will not arise/is addressable in the way we get to typical quantum theories of gravity. Take loop quantum gravity (LQG): LQG is constructed by a quantisation scheme from general relativity. Surely, there are ambiguities and thus choices to be made on the road—but at no point is it unintelligible how LQG arises *from the viewpoint of* GR. And that arguably addresses the worry that it is hard to link up the spatiotemporal to the non-spatiotemporal: we should be concerned with linking from the spatiotemporal to the non-spatiotemporal because it is the *spatiotemporal*, not the *non*-spatiotemporal, that we can understand.

# Acknowledgements

We are grateful to Chris Smeenk and to the audience of the Bonn work-in-progress seminar for helpful feedback.